\input psfig                                                       
\magnification=\magstep1
\font\naslov=cmbx10 %scaled \magstep1

\def\frac#1#2{{\begingroup#1\endgroup\over#2}}

\centerline{\naslov ON THE IONIZING SOURCES IN SPIRAL GALAXIES:}

\centerline{\naslov II. THE CENTER OF THE MILKY WAY}

\medskip

\centerline{V. MILO{\v S}EVI{\'C}--ZDJELAR$^{1*}$, 
S. SAMUROVI{\'C}$^{1**}$ AND  M.  M. {\'C}IRKOVI{\'C}$^{2,3***}$}

\medskip

\centerline{\it $^1$ Public Obs., Gornji Grad 16, Kalemegdan, 11000
Belgrade, Yugoslavia}

\centerline{\it $^{*}$ E-mail vesnamz@afrodita.rcub.bg.ac.yu}

\centerline{\it $^{**}$E-mail srdjanss@afrodita.rcub.bg.ac.yu}

\centerline{\it $^2$  SUNY at Stony Brook,  Stony Brook, NY 11794-3800, USA}

\centerline{\it $^3$ Astronomical Obs., Volgina 7, 11000 Belgrade,
Yugoslavia}

\centerline{\it $^{***}$E-mail  cirkovic@sbast3.ess.sunysb.edu}

%\medskip

\bigskip
\noindent{\bf 1. INTRODUCTION}
\medskip

\noindent The overall shape of our galaxy, Milky Way,  has mainly been revealed -- it is 
classified as a SAB(rs) type galaxy with a bar (Kuijken 1996, and references 
therein; Gyuk 1996), weak rings and a four arms spiral structure (Val\'ee 
1995). 

A model for the rotation curves of spiral galaxies by Sofue 
(Sofue 1996), could be applied to the Milky Way introducing:

%\medskip

\item{$\bullet$}  nuclear mass component (scale radius $ r \sim 100-150$ pc,$\;$ mass $M 
\sim 3-5 \times \ 10^9$ $M_{\odot}$),  

\item{$\bullet$}
  central bulge ($ r \sim 0.5-1$ kpc, $M \sim 10^{10}\; M_{\odot}$),

\item{$\bullet$}
  disk ($ r \sim 5-7$ kpc, thickness $\sim 0.5$ kpc ,  and $M \sim 1-2 \times 
10^{11}\; M_{\odot}$),

\item{$\bullet$}
  massive halo ($ r \sim 15-20$ kpc, $M \sim 2-3 \times 10^{11} \; M_{\odot}$).

The fifth component could be added (Sofue 1996) for the Milky Way -- the very 
nucleus (${ r \sim 30\; {\rm pc},  
M \sim 10^7 M_{\odot}}$) with a "dark" mass at a 
dynamical center of the Galaxy (${r \leq 0.01\; {\rm pc},    M \simeq 2.6 
\times 10^6 M_{\odot}}$). 

The distribution and influence of different  ionizing sources throughout the 
Galaxy (even some local like the great Gum Nebula) could  be described 
by the Taylor-Cordes model (TC93) (Taylor and Cordes 1993), with
Galactic center (GC)
 component 
added (Lazio and Cordes 1998, hereafter LC98), using a number density of free electrons 
in the interstellar medium as a parameter of ionization. In this contribution 
we discuss the fifth (GC) component that becomes dominant from 
the galactocentric distance of 0.5 kpc towards the GC. 

The opened question remains whether the "dark mass" in GC is directly 
responsible for the high ionization of that region, or only indirectly 
through the 
accretion disk, if it is a supermassive black hole (BH).

\bigskip
\noindent{\bf 2. THE CENTER OF THE MILKY WAY}
\medskip

\noindent The complete activity of the GC region has been studied for almost two 
decades now, and numerous nonthermal and thermal sources have been revealed.
All of them contribute to the total electron density of the last component in 
the equation that represents the overall estimate of electron density ($n_e$) 
distribution (TC93, LC98): 

\vskip-0.5cm

$$n_e(x,y,z)=n_1g_1(r){\rm sech}^2(z/h_1)+
n_2g_2(r){\rm sech}^2(z/h_2)+$$
\vskip-0.7cm
$$+ n_a{\rm sech}^2(z/h_a)\sum _{j=1}^4 f_jg_a(r,s_j)+
n_Gg_G(u)+
n_{\rm GC}g_{\rm GC}(r)h_{\rm GC}(z)\eqno(1)$$

%\medskip

\noindent where $r$ is the 
 Galactocentric
distance projected onto the plane and is equal $r=(x^2+y^2)^{1/2}$, the 
sum goes over four spiral arms, $n_i,\,  i=1\dots 5$ denotes the density
in different regions, $f_i,\, i=1\dots 4$ are scale factors,
 $g_i, \, i=1\dots 5$ are functions of position, $h_i,\; i=1\dots 4$ are
scale heights and $z$ is the height above the galactic plane. 
The detailed description of each component is given in TC93 
and LC98. 

The GC component has a following shape:

$$ n_{\rm GC}g_{\rm GC}(r)h_{\rm GC}(z)  =   
(10 \; {\rm cm}^{-3}) \times e^{-\left ({r \over 0.15} \right ){}^2} \times e^{-
\left ({z \over 0.075} \right ){}^2}\eqno (2)$$

The overall contribution of the GC component to the total electron density 
$n_e$ of the Galaxy, according to eq.~(2), is shown  in Fig. 1. The problem of 
the ionization throughout a spiral galaxy is briefly discussed in Samurovi\'c, 
\'Cirkovi\'c, Milo\v sevi\'c-Zdjelar and Petrovi\'c (1998) (Paper I).

Electron 
densities as  functions of $r$ are presented in TC93 (in their Fig.~3).              

Local contributions from the GC area come from a complex set of sources. At a  
dynamical center of the Milky Way lies a nonthermal synchrotron radio-source 
${\rm SgrA}^*$, of yet unrevealed nature, surrounded with thermal orbiting 
plasma SgrA West (within a Central Cavity,  $r \sim 1$ pc), which is ionized by 
IRS 16, a cluster of hot HeI/HI (spectral type O) stars in the vicinity of GC 
($r \sim 0.6$ pc) (Zylka {\it et al.}, 1995). Cavities like this one are usual 
features around stellar associations which blow out 
surrounding interstellar matter
by stellar winds, leaving hot rarefied gas cavity surrounded 
with envelope of neutral and ionized gas (Bochkarev and Ryabov 1997). In our 
case the envelope is a Circum-Nuclear Disc beginning at the outer edge of SgrA 
West (the Arc) at ($r \sim 1.7$ pc) and extending as far as 12 pc. SgrA West 
complex consists of a three arm minispiral and an extended ionized component, 
and contributes with $n_e \sim 10^4\; {\rm cm}^{-3}$ (minispiral) and $10^3 
\; {\rm \;
cm}^{-3}$ (extended component) (Beckert {\it et al.}, 1996). 

Some other local features of the smaller scale, like the Bullet (Yusef-Zadeh et 
al 1998), and the Sickle (Yusef-Zadeh {\it et al.}, 1997b),  contribute the $n_e \sim 
10^4 \; {\rm cm}^{-3}$ and $10^2\;  {\rm cm}^{-3}$, respectively. 

Another contribution of $n_e \sim 6\; {\rm cm}^{-3}$  (Koyama {\it et al.}, 1996) comes 
from a probable supernova remnant SgrA East located 30 pc behind the GC. It is 
heated by a cluster of hot O stars behind SgrA West (Sofue 1993). There have 
been some attempts to describe its unusual feature, more energetic than 
supernova, as a Seyfert-like activity like in the nuclei of some other spiral 
galaxies (LaRosa and Kassim 1985). SgrA East is physically interacting with a  
"50 km ${\rm s}^{-1}$ molecular cloud" forming new stars in the areas of 
collision, and is considered to be the source of the high energy activity of 
the GC (Yusef-Zadeh 1997a). The GC region ($r\sim$ few hundred pc) is 
responsible for 10\% of the total star forming rate of the entire Galaxy 
(Sofue 1993). 

The contribution of SgrA complex to the total electron density is $n_e \sim 6 
\; {\rm cm}^{-3}$, and it drops outside of that area to 0.3-0.4 ${\rm cm}^{-3}$.

One of the largest luminous HII plasma/molecular region in the GC vicinity is
 SgrB2 at a distance of 100 pc from GC. It consists of several active star 
forming regions with numerous associated dense HII regions (Gordon {\it et al.}, 
1993). Besides SgrB2, within the Nuclear Disc ($r \sim$ 200 pc, thickness 50 
pc), there are several other HII (SgrC, D, and E), and star forming regions.

ASCA 
observations detected strong K$\alpha$ lines from highly ionized various 
elements, and showed presence of high temperature plasma over the GC region 
(Koyama {\it et al.}, 1996).

On the large scale, hot plasma is distributed symmetrically along the galactic 
plane with a strong concentration at the GC. X-ray spectra obtained from ASCA 
showed high energy (10 keV) spectra of the similar shape over the emission 
region (Koyama {\it et al.}, 1996). Their results show that energy generation rate 
resembles the rates at the active galactic (Seyfert-like) nuclei, having a 
large mass concentration at the GC.

Weaker thermal emission is detected extending 80 pc along the Galactic plane 
on both sides of GC. Vertical to the Galactic plane, various thermal arched 
filaments extending more than 100 pc, are detected along magnetic field, 
having turbulent motion, preferably towards the GC. Similar 
vertical structures exist in other galaxies (Sofue 1993).

\bigskip
\noindent{\bf 3. THE NATURE OF Sgr$A^*$}
\medskip

\noindent Motions of the ionized gas can not reveal characteristics of the central 
object as the stars can, because the gas is influenced by the magnetic field. 
Also, high temperature plasma can not be bound by the galactic gravity.  

By proper motions of the stars within central 0.01 pc one can draw conclusions 
about the mass and nature of the central object (Yusef-Zadeh 1998). The 
most recent research  (Eckart and Genzel 1997) finds the value of stellar 
proper motion greater than 1000 km ${\rm s}^{-1}$ and undoubtedly estimates 
the central mass: 2.6 $\times 10^6 \; {M}_{\odot}$. 

Different models could be applied to the Sgr$A^*$ -- a central "dark mass".
If we consider a black hole model, we have to examine all the problems related 
to such an object. Radio-emission detected from GC due to cyclo-synchrotron 
and synchrotron radiation is consistent with a theoretical prediction of an 
accreting disk around a $10^6 \; {\rm M}_{\odot}$ BH (Bower and Backer 1998). 
There are few models successfully applicable to it:

\item{$\bullet$} advection dominated accretion (Narayan {\it et al.}, 1995,
Lasota 1998)

\item{$\bullet$}
 spherical accretion (Melia 1994).

The crucial problem related to the BH model is low X and $\gamma$-ray 
flux detected from Sgr$A^*$. The strongest source in the vicinity of GC is 1E 
1740.7-1942, hard X-ray source, and a strong source of annihilation 511 keV 
line (e.g. Wehrse {\it et al.}, 1996), but it is not coinciding with  Sgr$A^*$ (it 
is situated in a dense molecular cloud 50' away from GC). It shows all the 
features (similar shape and luminosity) as CygX-1, another BH in a 
Milky Way.

There have recently been some quite different approaches to the "blackness" of 
the Sgr$A^*$.
We briefly  mention here the recent proposal that Sgr$A^*$ is a neutrino ball,
 i.e. there is no BH  -- instead there is 
 a ball made of self-gravitating, degenerate
 neutrinos with the same total mass of $2.5 \times 10^6\; 
{M}_\odot$ (Tsiklauri 
and Viollier 1998a,b). These neutrinos have masses $m_\nu\ge 12$ keV/$c^2$
(for $g=2$) and $m_\nu\ge 14.3$ keV/$c^2$ (for $g=1$), where $g$ is the spin 
degeneracy factor. These neutrinos are, according to cosmological constraints,
 decaying thus producing X-ray emission lines. 
Another way to distinguish the existence of the neutrino ball is to
examine in detail the trajectory of stars in the vicinity of the GC. 
If there is 
the BH the trajectory will be  an ellipse with the BH at the focus, while 
in the case of the neutrino ball the center of the ellipse will be 
the center of the ball.

\bigskip
\vfill\eject
\noindent {\bf REFERENCES}
\smallskip

\item{}\kern-\parindent{Beckert, T.,   Duschl, W.J., 
  Mezger, P.J. and  Zylka, R.: 1996, 
{\it Astron. Astrophys.}, {\bf 307}, 450.}

\item{}\kern-\parindent{Bochkarev N., and Ryabov, M.: 1997 {\it Astrophys. 
Space Sci.} {\bf 252}, 309.}

\item{}\kern-\parindent{Bower, G.C. and  Backer, D.C.: 1998, {\it Astrophys. 
J.} {\bf 496}, 97.}

\item{}\kern-\parindent{Eckart, A. and Genzel, R.: 1997, {\it MNRAS}, {\bf 284}, 576.}

\item{}\kern-\parindent{Gordon, M.A., Berkermann, U., Mezger, P.J., 
 Zylka, R., 
 Haslam, C.G.T.,  Kreysa, E.,  Sievers, A. and  Lemke, R.:
1993,  {\it Astron. Astrophys.}, {\bf 280},  208.}

\item{}\kern-\parindent{Gyuk G.: 1996, preprint astro-ph/9607134.}

\item{}\kern-\parindent{Koyama, K., Maeda, Y., Sonobe, T., Takeshima, T., 
Tanaka, Y., Yamauchi, S.: 1996, {\it PASJ}, {\bf 48},  249.}

\item{}\kern-\parindent{Kuijken, K.: 1996, in {\it IAU Coll. 157}, eds. R. 
Buta,  D.A. Crocker, B.G. Elmegreen, ASP, San Francisco, 504.}

\item{}\kern-\parindent{LaRosa, T., N., Kassim, N., E.: 1985, 
{\it Astrophys. J.}, {\bf 299}, L13.}      

\item{}\kern-\parindent{Lasota, J.-P.: 1998, {\it Phys. Rep.} (to be
published), preprint astro-ph/9806064.}

\item{}\kern-\parindent{Lazio, T.J.W. and  Cordes, J.M.: 1998 {\it Astrophys. 
J.}, {\bf 497},  238 (LC98).}

\item{}\kern-\parindent{Melia, F.: 1994, {\it Astrophys. 
J.}, {\bf 426}, 577.} 

\item{}\kern-\parindent{Samurovi\'c, 
S., \'Cirkovi\'c, M.M.,  Milo\v sevi\'c-Zdjelar, V. and Petrovi\'c, J.: 1998,
{\it Proceedings of 19th Symposium of the Physics of the Ionized Gases}, 
submitted (Paper I).}

\item{}\kern-\parindent{Sofue, Y.: 1993, NATO adv. school "The Nuclei of 
Normal Galaxies: Lessons from the Galactic Center", Ringberg Schloss 
Tegernsee, 
preprint astro-ph/9309046.}

\item{}\kern-\parindent{Sofue, Y.: 1996, {\it Astrophys. J.}, {\bf 458}, 120.}

\item{}\kern-\parindent{Taylor, J.H. and  Cordes J.M.: 1993 {\it Astrophys. 
J.} {\bf 411}, 674 (TC93).}

\item{}\kern-\parindent{Tsiklauri, D. and Viollier, R.D.: 1998a, {\it 
Astrophys. J.}, submitted (preprint 

\ \ \ \ astro-ph/9805272).} 

\item{}\kern-\parindent{Tsiklauri, D. and Viollier, R.D.: 1998b, {\it 
Astrophys. J.}, {\bf 500}, 591 (preprint\ \ \  \ astro-ph/9805273).}

\item{}\kern-\parindent{Val\'ee, J.P.: 1995, {\it Astrophys. J.} {\bf 454}, 
119.}

\item{}\kern-\parindent{Wehrse, R., Duschl, W. J., Hof M., and Tscharnuter, W. 
M.: 1996, {\it Astron. Astrophys.}, {\bf 313}, 457.}

\item{}\kern-\parindent{Yusef-Zadeh, F.,  Purcell W. and  Gotthelf, E.: 1997a,
 {\it 
IAUS} {\bf 184}, 196.}

\item{}\kern-\parindent{Yusef-Zadeh, F.,   Roberts, D.A. 
 and  Wardle, M.: 1997b,  {\it 
Astrophys. J.} {\bf 490}, 83.}

\item{}\kern-\parindent{Yusef-Zadeh, F.,  Roberts, D.A. and  Biretta, J.: 
1998,  {\it 
Astrophys. J.}, {\bf 499}, 159.}

\item{}\kern-\parindent{Zylka, R., Mezger P.G.,  Ward-Thompson, D.,  
Duschl, W. J. and  Lesch, H.: 1995, {\it Astron. Astrophys.}, {\bf 297}, 83.}

\vskip20cm
\psfig{file=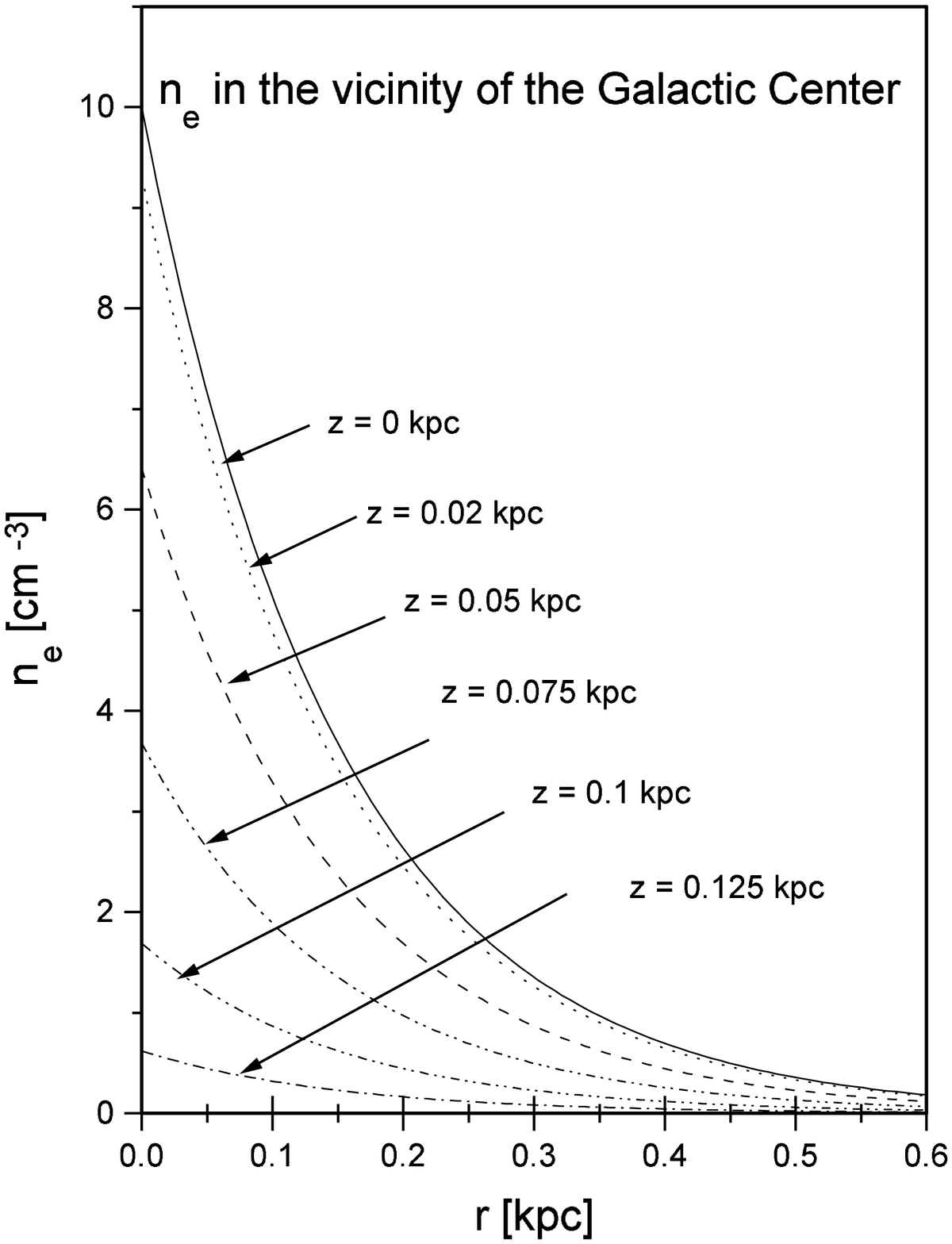,height=14truecm}

\bigskip
\bigskip

\noindent Figure 1.
Dependence of electron density ${n_e}$
 as a function of 
Galactocentric radius ${r}$
 in the vicinity of the GC (up to ${\sim}$ 500 pc) for 
different heights above the galactic plane -- ${z}$, according to LC98. 

\bye